\documentclass[sigconf]{acmart}
 \onecolumn
\AtBeginDocument{%
  \providecommand\BibTeX{{%
    \normalfont B\kern-0.5em{\scshape i\kern-0.25em b}\kern-0.8em\TeX}}}

\setcopyright{acmcopyright}
\copyrightyear{2020}
\acmYear{2020}
\acmDOI{10.1145/1122445.1122456}

\usepackage{hhline}
\usepackage{subcaption}

\begin{document}

\title{Hardware-accelerated Simulation-based Inference of Stochastic Epidemiology Models for COVID-19}

\author{Sourabh Kulkarni}
\email{skulkarni@umass.edu}
\affiliation{%
  \institution{University of Massachusetts Amherst}
  \streetaddress{151 Holdsworth Way}
  \city{Amherst}
  \state{Massachusetts}
  \country{USA}
  \postcode{01003}
}

\author{Mario Michael Krell}
\affiliation{%
 \institution{Graphcore}
 \streetaddress{167 Hamilton Ave, Palo Alto, CA 94301}
 \city{Palo Alto}
 \state{CA}
 \country{USA}}

\author{Seth Nabarro}
\affiliation{%
  \institution{Graphcore Research}
  \streetaddress{Prudential Buildings Wine St, Bristol BS1 2PH}
  \city{Bristol}
  \country{UK}}

% \author{Alex}
% \affiliation{%
%   \institution{Graphcore}

\author{Csaba Andras Moritz}
\affiliation{%
  \institution{University of Massachusetts Amherst}
  \streetaddress{151 Holdsworth Way}
  \city{Amherst}
  \state{Massachusetts}
  \country{USA}}

\renewcommand{\shortauthors}{Kulkarni, et al.}

\begin{abstract}
Epidemiology models are central in understanding and controlling large scale pandemics. Several epidemiology models require simulation-based inference such as Approximate Bayesian Computation (ABC) to fit their parameters to observations. ABC inference is highly amenable to efficient hardware acceleration. In this work, we develop parallel ABC inference of a stochastic epidemiology model for COVID-19. The statistical inference framework is implemented and compared on Intel Xeon CPU, NVIDIA Tesla V100 GPU and the Graphcore Mk1 IPU, and the results are discussed in the context of their computational architectures. Results show that GPUs are 4x and IPUs are 30x faster than Xeon CPUs. Extensive performance analysis indicates that the difference between IPU and GPU can be attributed to higher communication bandwidth, closeness of memory to compute, and higher compute power in the IPU. The proposed framework scales across $16$ IPUs, with scaling overhead not exceeding $8\%$ for the experiments performed. We present an example of our framework in practice, performing inference on the epidemiology model across three countries, and giving a brief overview of the results.
\end{abstract}




\maketitle
 \onecolumn
\section{Introduction}

The key objective of epidemiology modelling is to capture the underlying mechanism of the spread of a disease in the population. 
These models help explain the current and past behaviour of the disease, 
as well as providing projections into the future. 
There exists a wide variety of models developed with different levels of complexity and incorporating a variety of data sources. In all cases, it is crucial to ``fit'' the model to the current disease outbreak. This involves inferring the key parameters such as infection rate, mortality rate, recovery rate, reproduction rate etc, from available observations. 

Bayesian inference has desirable properties in this context, namely the incorporation of prior belief and the ability to infer a distribution over  parameters rather than predicting a point value. Exact Bayesian inference is tractable only for a small set of parametric distributions. Traditionally, epidemiological modelling has circumvented this with approximate Bayesian inference methods belonging to the Markov Chain Monte Carlo (MCMC) family \cite{hamra2013markov}. While these inference methods are effective for many models, their compatibility for a given model depends of the tractability of the \textit{likelihood function} - the ability to compute probability of observed data for a given set of parameters. This requirement is not satisfied for several epidemiological models, due of presence of unobserved or ``latent" sub-populations, for example the unconfirmed infections. 
Due to this incompatibility, either the development of epidemiology models has to be restricted to a sub-class where all sub-populations are observed, or a new statistical inference method is to be developed which does not require the computation of the likelihood function for parameter inference. 
Over the last decade several such algorithms have been 
developed~\cite{abc_evolution,abc_evolution2,abc_cosmology,abc_econometrics,abc_cognitive,abc_sysbio,abc_biochem}, 
which are together grouped into the class of ``likelihood-free inference'', 
also known as simulation-based inference. 
At their core, these algorithms exploit the ability to simulate the model to perform inference over it. Such approaches have been successfully applied to a variety of domains including evolution and 
ecology~\cite{abc_evolution,abc_evolution2}, cosmology~\cite{abc_cosmology}, 
econometrics~\cite{abc_econometrics}, cognitive science~\cite{abc_cognitive}, 
systems biology~\cite{abc_sysbio}, and biochemistry~\cite{abc_biochem}.

In this work, we consider one such simulation-based inference method -- Approximate Bayesian Computation (ABC) \cite{rubin1984bayesianly}. 
We develop a parallelized version of ABC which allows us 
to perform highly efficient inference by leveraging recent machine-learning optimized hardware, and software libraries \cite{tensorflow2015-whitepaper}. 
This version of ABC is used for accelerated inference of the stochastic epidemiological model presented in \cite{Warne2020},
which is applied to help understanding and predicting the spread of COVID-19 using real-world data. We note the original implementation~\cite{Warne2020_github} was run on CPUs in HPC clusters, and so did not make use of specialist hardware acceleration. There has been some work in hardware acceleration and distributed computing for ABC \cite{liepe2010abc,dutta2017abcpy}, however there is, to our knowledge, no prior work which i) systematically analyzes and attempts to understand the performance of different hardware accelerators for the proposed parallel version of ABC and ii) demonstrates hardware acceleration of ABC for epidemiology modelling. We were thus motivated to investigate how effective the NVIDIA Tesla V100 GPU and Graphcore Mk1 IPU are in accelerating the inference algorithm in question.

We observe that the parallelized ABC inference over the model performs $\approx 4\times$ and $\approx 30\times$ faster on the Tesla V100 GPU and  Mk1 IPU vs. the Xeon Gold CPU. To better understand the performance benefits with the hardware acceleration platforms, we do an in-depth analysis with different algorithmic configurations. This analysis includes batch-size sweeps, communications overhead and host processing, memory utilization, and processing load distribution. Our analysis indicates three key insights on how the IPU architecture performs $\approx 7\times$ better than the GPU architecture for this workload: i) The much larger on-chip cache (300MB on IPU vs. 6MB L1 + 10MB L2 on V100 GPU), making the IPUs scale better in performance with increasing batch sizes;  ii) The overhead in GPU of fetching execution code from the main memory to the compute units vs. no such overhead in IPUs as code is already `at' the compute unit; and iii) The much faster memory bandwidth of IPU ($45$TB/s) vs. V100 GPU ($800$GB/s) and much higher compute power ($62$ TFLOPS vs. $14$ in GPU) which leads to much faster overall computation. 

To further investigate the scalability of the proposed parallel ABC inference approach, we demonstrate how the runtime of the inference varies with varying compute resource, running on between $2$ to $16$ IPU devices. We observe a near linear reduction in runtime with number of devices, seeing a minimal scaling overhead of up to $8\%$ depending on the configuration, and how much sample post-processing is done on-device. 

We demonstrate our model in practice by performing a comparative analysis of the epidemiology model across three countries (Italy, USA and New Zealand). We plot predictions of the model with parameters inferred from real-world data, and analyze the parameters to gain insights over the responses of these countries to the epidemic. The analysis for each country took $\approx 1.9$ hours on a 16-IPU system, while we estimate $\approx 57$ hours on $16$ Xeon CPUs and $\approx 15$ hours runtime on $8$ Tesla V100s for the same analysis. These results demonstrate that massively parallel simulation-based inference can be efficiently enabled with hardware acceleration, paving the way for epidemiology modelling with faster results and better sample estimates.

The paper is organized as follows: Section~\ref{s:background} provides background information about the epidemiology models and simulation-based inference algorithms. 
Section~\ref{sec:design} walks though the design methodology for a high-performance implementation of the simulation-based inference for stochastic epidemiology models. 
Section~\ref{s:performance} provides detailed performance analysis of two hardware acceleration platforms 
over which this proposed algorithm is executed. 
Section~\ref{s:analysis} discusses results of running the model on real-world COVID-19 data. 
Section~\ref{s:conc} concludes the paper. 

\section{Background}
\label{s:background}

% - overview of epidemology models SIR, etc

% - notes on key similarities and differences among epidemology models of this class (SIR type)

% - overview of the specific model used, math desc etc.

% - simulation based inference overview

% - ABC overview and key similarities and differences in ABC and other simulation-based inference methods

% - here we can make a case that the 'simulation' part remains same across different algorithms, hence many aspects of the performance analysis we preform are valid across several simulation-based inference algos.

In this section, we provide a brief introduction and mathematical description of the Stochastic Epidemiology Model being considered.
We also provide details on the ABC inference algorithm. 
Finally, we discuss the hardware acceleration platforms used in this work. 

First, we shall establish the terminology used across the rest of the paper. 
A model is represented as a joint probability distribution $p(\theta, x)$ where $\theta$ is the set of model parameters, 
on which we wish to perform inference, and $x$ denotes the observed variables of the model, which can then be compared to real-world data.
The parameters are typically believed to lie in a certain distribution, often based on domain expertise. 
This information is encoded in the prior over those parameters, denoted by $\pi = p(\theta)$.
The model with $\theta$ specified can be used to generate observations over $x$; this process is called model simulation and the simulated data is referred to as $D_s$. 
For a stochastic simulator, the expression $D_s \sim p(x|\theta)$ concisely represents this simulation process.
The prior belief over parameters distributions $p(\theta)$ can be updated by conditioning on observed real-world data $D$. 
This updated parameter distribution is known as the posterior, 
and is denoted by $p(\theta|x=D)$, or simply $p(\theta|D)$. 
The likelihood function, 
which represents the probability of observing the real-world data $D$ given our parameter setting $\theta$, 
is denoted by $p(x=D|\theta)$, or $p(D|\theta)$.

\subsection{Stochastic Epidemiology Model}
\label{s:epi}
\begin{figure*}[htbp!]
\centerline{\includegraphics[width=\linewidth]{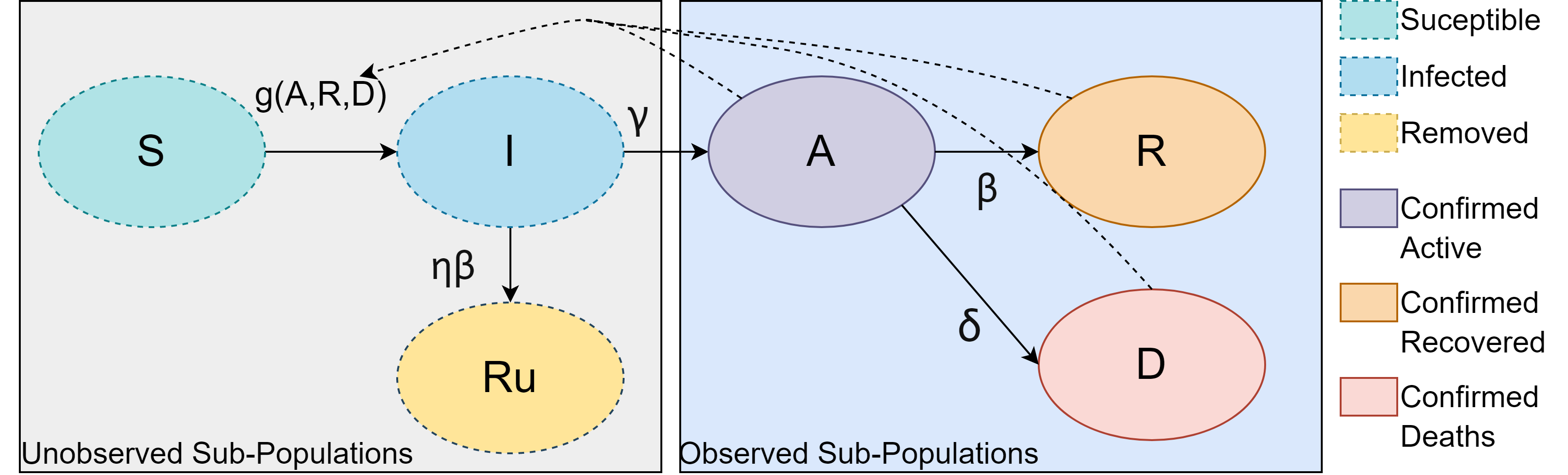}}
\caption{Overview of the epidemiology model flow. The population is divided into $6$ sub-populations. 
On a per-day basis, the number of transitions from one sub-population to the other are simulated with a Poisson process, where the rates are governed by the current sub-populations and the transition parameters. 
The transition from susceptible to infected is a function of the observed sub-populations 
which captures the response of a population to an increasing number of cases. 
}
\label{f:model_overview}
\end{figure*}

The epidemiology model considered in this work belongs to the class known as compartmental models. In this class of models, the population is divided into sub-populations, and the spread of infectious disease is modelled as the flow from one sub-population to other. 
The models in this class vary in number of sub-populations, and how the flow among them is described. 
The most basic version of this model class is the SIR model (Susceptible, Infected, Recovered)~\cite{SIR_paper}, which attempts to capture the dynamics of a contagion in those three sub-populations. 
There have been several models of this class that build on this basic framework by adding more sub-populations and new methods of modelling flow among these sub-populations~\cite{DiffusionReactionModel,StochasticSemiMarkov}.
In recent years models have also added stochastic dynamics to capture the inherent probabilistic nature of spread of disease in a population~\cite{StochasticSemiMarkov,StochasticEbola,BelgiumStochasticCOVID}. 
Some also classify sub-populations as either being observed or latent, 
to capture the notion of having sub-populations of untested individuals~\cite{LatentSEIR,Warne2020}.

The model considered in this work~\cite{Warne2020} attempts to capture the spread of COVID-19 
using six sub-populations, three of which are observed and three unobserved. We include a brief overview of the model, but refer the reader to the Supplementary Material of \cite{Warne2020} for further detail.
The transmission across these sub-populations is modeled with a Poisson process approximated in discrete timebins using the tau-leaping method \cite{tau_leaping}. 
Fig.~\ref{f:model_overview} provides an overview of the model. 
The model consists of $8$ parameters:

\begin{equation}
    \theta = [\alpha_0, \alpha, n, \beta, \gamma, \delta, \eta, \kappa]
\label{parameters}
\end{equation}

with a uniform prior distribution:

\begin{equation}
    \pi = p(\theta) = U(0, [1, 100, 2, 1, 1, 1, 1, 2])
\end{equation}

These prior values were taken as-is from the original model description~\cite{Warne2020}.
They signify the reasonable ranges in which the parameters of interest could lie. 
Simulating the model with a sample from parameter distributions provides the following state vector:

\begin{equation}
X = [S, I, A, R, D, R^u],
\end{equation}
which consists of the sub-populations of
\textbf{S}usceptible people, undocumented \textbf{I}nfected, 
\textbf{A}ctive confirmed cases, confirmed \textbf{R}ecoveries,
confirmed fatalities \textbf{D}ying,
and unconfirmed \textbf{R}$^u$emoved. The Removed sub-population, $R^u$, comprises those who have recovered or died, but have not been tested. This simulation is typically performed over several days or months and the generated data can be compared with the real-world values of the observable sub-populations.

One of the key challenges of this model is that $X$ is \textit{partially} observed; i.e. only the $A, R, D$ values are available from observed data.
This makes the likelihood function $p(D|\theta)$ intractable for this model, as the unobserved sub-populations of the model are required to be `integrated-out'.
Instead, simulation-based inference such as ABC is used to perform inference over this model.

Parameter $\alpha_0$ refers to the base infection rate, while $\alpha$ is the coefficient of the function that captures the changes in infection rate based on the observed sub-populations (A,R,D). $n$ is the exponent to the function. Based on these parameters the total infection rate is assumed to follow:

\begin{equation}
    g_{(A,R,D)} = \alpha_0 + \frac{\alpha}{1 + (A+R+D)^n}
\label{total_infection_rate}
\end{equation}

This function can be modified to capture additional changes to infection rate based on $A$, $R$, $D$ values or even using external data.

The parameters $\gamma$, $\beta$, and $\delta$ are the positive test rate, recovery rate and fatality rate respectively.  
The parameter $\eta$ captures the effectiveness of testing protocols, 
as the rate for unconfirmed infected to be recovered without ever being confirmed is given by $\eta\beta$. 
The initial value parameter, $\kappa$, encodes the number of unobserved infected cases, as a fraction of $A$, at the start of the simulation.

The underlying COVID-19 time-series data, 
provided by Johns Hopkins University~\cite{Dong2020},
contains daily numbers for $[A, R, D]$.

In its first step, the model initializes the remaining variables
with $R^u=0$, $I_0 = \kappa * A_0$, and 
$S=P-(A_0+R_0+D_0+I_0)$ with $P$ being the total population count at
the first time point.

The second step is to calculate the hazard function $h$
which provides the average update in the model parameters
within one day

\begin{equation}
    h(S, I, A, R, D, R^u)=
    \left(g S\frac{I}{P},
    \gamma  I,
    \beta  A,
    \delta  A,
    \beta \eta I
    \right).
\end{equation}
with $g$ described in equation \ref{total_infection_rate}.

The third step is to randomly sample the transmission amounts according to these average numbers.
Instead of a Poisson sampling with $h$ as parameter,
we chose an approximation with 
normal distributions with mean $h$ and variance $\sqrt{h}$
and use the floor of the numbers.

The fourth step is to apply the sampled transmission amounts to obtain updated
numbers for the next day 
($S\rightarrow I$, $I\rightarrow A$, $A\rightarrow R$, $A\rightarrow D$,
$I\rightarrow R^u$, ordering according to $h$ function).

The second to fourth step are repeated in a loop for each day. 
Eventually, the numbers for $A$, $R$, and $D$ can be compared to the real
measurements.

\subsection{Approximate Bayesian Computation}
\label{s:ABC}
The standard Bayesian statistical inference approach of obtaining the posterior over parameters $\theta$ for a model $p(\theta, x)$ and given observations $D$, is given by Bayes' rule,

\begin{equation}
    p(\theta|D) = \frac{p(D|\theta)p(\theta)}{p(D)} 
\end{equation}
which describes how to update the belief about our parameters $\theta$ given observations $D$. As discussed earlier, the likelihood function is intractable, since $S_t$, $I_t$, and $R^u_t$ are unobserved. This precludes some approximate Bayesian inference methods such as MCMC.
In the ABC approach, the ability to simulate from the model is utilized to perform inference on it. 
First, we sample the parameters $\theta$ from their prior $\theta^* \sim \pi$.
Next, we simulate a forward pass of the simulator (as described in Section~\ref{s:epi} for example) to generate observations $D_s \sim p(x|\theta^*)$ over the number of days we have data for. 
The simulated observations are then compared to the real-world evidence using a distance function $dist(D_s, D)$. 
For this model we used the Euclidean distance~\cite{Warne2020}. 
Finally, the sampled parameters $\theta^*$ are accepted 
if the distance function is less than a certain tolerance value 
$\epsilon$, $dist(D_s, D) \leq \epsilon$. 
This is repeated until we accept the target number of posterior samples. 
In essence this ABC process is sampling parameters from the approximate posterior of the model given the data while effectively circumventing the likelihood function. 
It can be shown that as tolerance $\epsilon$ approaches $0$, the approximate posterior converges to the true posterior~\cite{ABC_paper}.

To summarize, in ABC we aim to obtain samples from an approximation to the posterior:
\begin{equation}
    p(\theta|D) \approx p(\theta|\text{dist}(D,D_s)\leq\epsilon) \propto 
    p(\text{dist}(D,D_s)\leq\epsilon | \theta)p(\theta)
\end{equation}
where $D$ is the ground truth data, $D_s$ is simulated data depending on $\theta$, and $p(\theta)$ is the prior~\cite{Warne2020}.
The $\text{dist}$ function is the Euclidean distance~\cite{Warne2020}.
An example for the distribution
$p(\text{dist}(D,D_s)\leq\epsilon | \theta)$
is provided in Figure~\ref{f:histograms1000}.
In our experiments, 
a uniform distribution is used for $p(\theta)$.

Instead of choosing a fixed threshold, sequential Monte Carlo can be 
used to transform an initial set of samples
to a high quality set with a decreasing sequence of thresholds $\epsilon$
and using ABC. This algorithm is called SMC-ABC~\cite{Warne2020, Drovandi2011}.

It is important to note that these statistical `inference' algorithms are solving a `learning' problem, and the expression `parameter inference' in the statistics literature corresponds to `parameter learning' in ML literature. This should not be confused with `inference' as used in the ML literature, which often means the process of estimating an output with a model, given the input. 

\subsection{Hardware Acceleration Platforms}
Recent advances in deep learning have driven the development of novel machine-learning hardware acceleration platforms. Improvements in computational capabilities of these platforms are an important factor in enabling bigger, more complex deep networks to be developed and deployed. While primarily focused on acceleration of deep-learning type workloads, these platforms could be potentially utilized for accelerating certain other machine-learning tasks. 

The hardware-accelerated simulation-based inference framework developed in this work is targeted towards two such platforms - the Nvidia Tesla V100 GPU and the Graphcore Mk1 IPU. In addition, we compare both to a Xeon Gold 6248 CPU baseline. 

\subsubsection{Tesla V100 GPU}
The current widely used hardware acceleration platform for several AI applications is the Tesla V100. 
It consists of $640$ Tensor cores and $5120$ CUDA cores. While each CUDA core can perform a single-precision (FP32) multiply-accumulate operation per clock-cycle, each Tensor core can perform a $4\times4$ multiply-accumulate operation per clock-cycle. 
 The solution has a reported $112$ TFLOPS of tensor performance, $14$ TFLOPS single precision,
and $900$ GB/s memory bandwidth, with a Thermal Design Power (TDP, peak power) of $300$W.\footnote{https://www.nvidia.com/en-us/data-center/v100/} The memory of the GPU is arranged in a hierarchy. It includes $80$ streaming multiprocessors (SMs), each with an L1 cache of $128$KB and therefore a total L1 cache size of $10$MB. The L2 cache is $6$MB and the main memory is $16$GB, though we find that only 14.38GB is accessible in practice. The GPU has a base clock of $1230$Mhz and a boost clock of $1380$Mhz.\footnote{https://images.nvidia.com/content/volta-architecture/pdf/volta-architecture-whitepaper.pdf}

The Tesla V100 has shown good 
performance in several benchmarks such as
MLPerf in training as well as inference~\cite{mlperf}.
Given its widespread use, 
this was chosen as one of the targets for exploring hardware-accelerated simulation-based inference. Our GPU implementation is written in TensorFlow 2.1 \cite{tensorflow2015-whitepaper}.

\subsubsection{Intelligence Processing Unit (IPU)}

The IPU is a MIMD (multiple instruction, multiple data) processor. 
It is well suited for problems that require fine-grain parallelism 
and high-speed memory access. 
In addition, the MIMD thread-level parallelism also benefits in models that use auto-regressive or sequential elements, 
employ random memory access patterns, 
or consist of non-vectorizable parallel paths.

A Mk1 IPU processor contains 1216 cores or ``tiles", each with its own local memory. A tile can run six parallel threads, thus a Mk1 IPU can run up to $1216\times6=7,296$ independent parallel threads. The bandwidth between the compute and the memory on the chip is $45$ TB/s. While the memory is very high bandwidth, the total memory size is much smaller than a GPU, with $300$MB on each IPU processor.
IPU-Links are used to connect multiple IPUs together for model-parallel and data-parallel execution.
Each Mk1 IPU offers an ``arithmetic throughput [of] up to $31.1$ TFlops/s 
in single precision and $124.5$ TFlops/s in mixed precision per chip''~\cite{IPU}. A C2 PCIe accelerator card, comprising two Mk1 IPUs has a TDP of $300$W\footnote{\url{https://www.graphcore.ai/hubfs/assets/pdf/C2\%20Card\%20Product\%20Brief.pdf?hsLang=en}}, the same as a single Tesla V100 GPU. The IPUs in the C2 card have a clock-speed of $1300$Mhz which is also comparable to the Tesla V100.
Hence, for most evaluations, we compare the performance of two IPUs against a single GPU.

There are three motivations to analyze IPU performance on ABC.
First, currently large amounts of CPUs are often used for processing ABC workloads~\cite{Warne2020_github}.
Taking advantage of the independent parallel processes on the IPU could
drastically reduce energy consumption and increase performance.
Note that this advantage can be easily leveraged with
a framework like TensorFlow.
Second, the simulation and ABC algorithm can fit in the SRAM memory
of the Mk1 IPU for reasonable batch sizes, which drastically reduces the communication overhead
and enables efficient in-processor computation.
Last, like the epidemiology model considered in this work, and unlike traditional neural-network models, there are several 
new applications with rather complicated
computational graphs comprising a large number of heterogeneous operations, rather than large scale matrix multiplications.
The IPU has been utilized in several such applications.
Examples are
natural language processing~\cite{bert_ipu}, image processing with modularized architectures~\cite{qwant}, 
bundle adjustment~\cite{bundle},
as well as some microbenchmarks~\cite{IPU}.

For the IPU evaluation, we have used a machine with $16$ Mk1 IPU processors ($8$ C2 cards). 
However, we run most experiments on only one of the eight cards. For implementation, we used the 1.2 version of 
its SDK with its IPU interface to TensorFlow 2.1 \cite{tensorflow2015-whitepaper}.

%We believe the high FLOPS, local memory paradigm of the IPU, and the lesser FLOPS, large memory architecture of the GPU present an interesting contrast.
We believe that the higher FLOPS and local memory paradigm of the IPU
presents and interesting contrast to the 
lesser FLOPS and large memory paradigm of the GPU. 
We seek to understand how each performs in the context of parallel ABC inference.

\section{Design Methodology}
\label{sec:design}

\subsection{Parallelized ABC Inference}
\label{s:model}

\begin{figure}[htbp]
\centering
     \begin{subfigure}[b]{0.47\textwidth}
        \includegraphics[width=\textwidth]{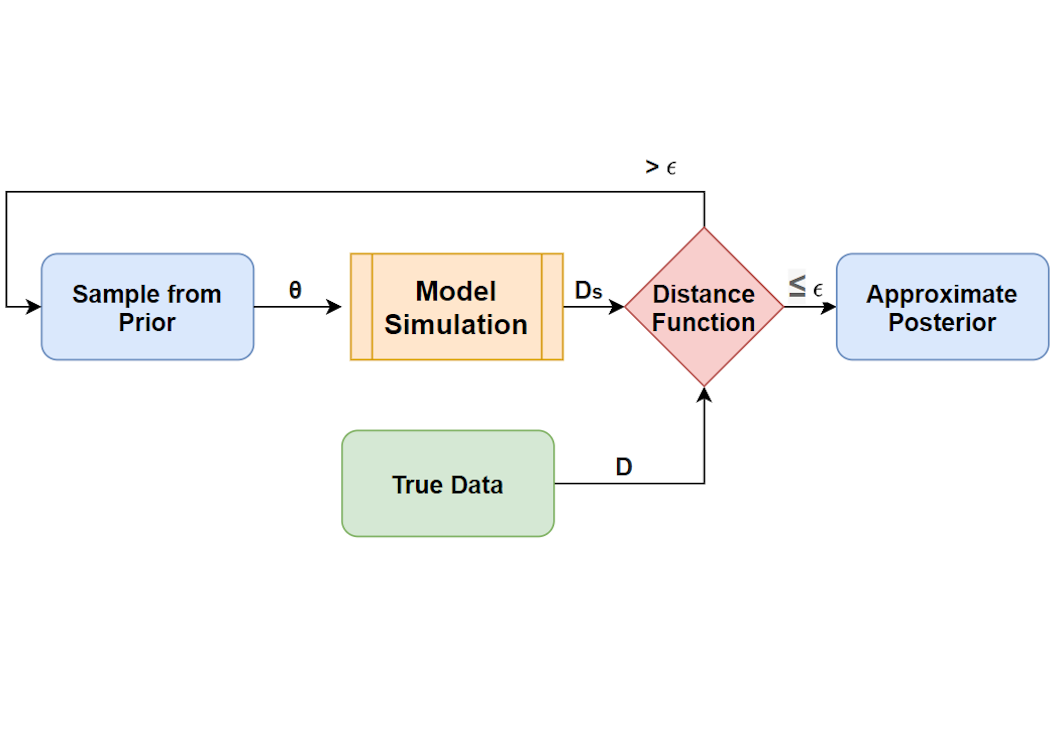}
         \caption{Vanilla ABC}
         \label{f:vanilla}
     \end{subfigure}
     \hspace{15pt}
    \begin{subfigure}[b]{0.47\textwidth}
        \includegraphics[width=\textwidth]{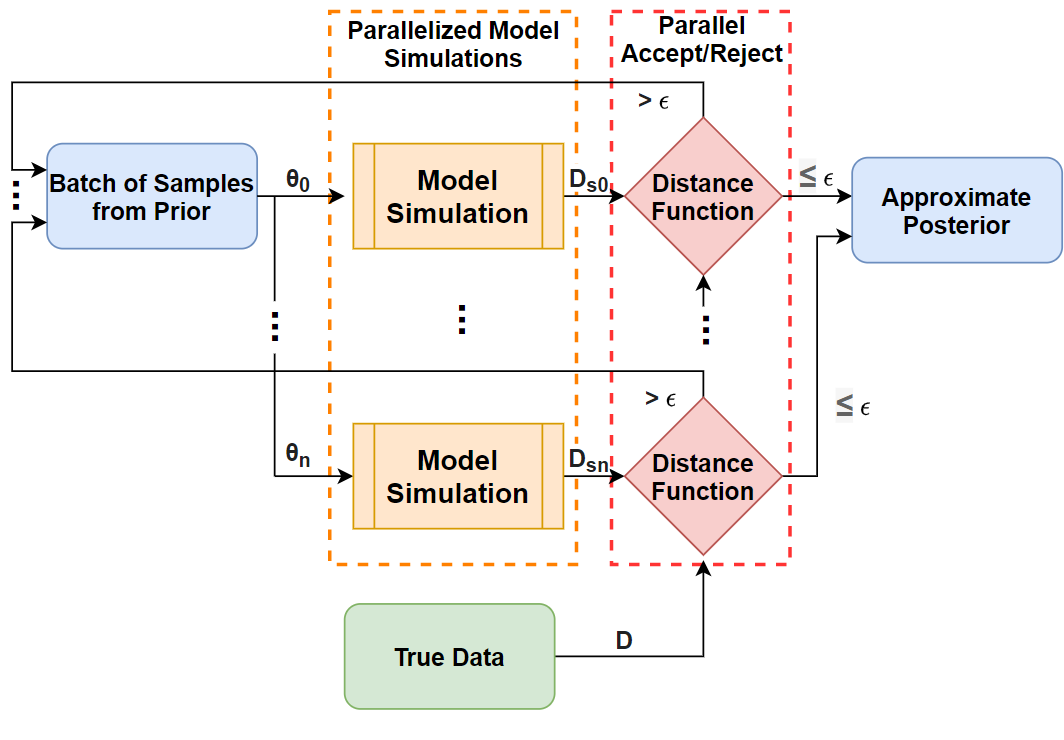}
        \caption{Parallel ABC}
        \label{f:vectorized}
     \end{subfigure}
\caption{A conceptual representation of parallelizing ABC inference. The sampling from prior, model simulation and distance computation are all embarrassingly parallelized, enabling support for hardware acceleration.}
\label{f:abc_parallelized}
\end{figure}

In the ABC inference process described in Sections~\ref{s:epi} and \ref{s:ABC},
the computational flow of sampling the parameters, simulating data and computing the distance function can be performed independently for any number of parameter samples. 
This is because the specific considerations usually involved in sequential Bayesian statistical inference algorithms like MCMC, such as burn-in, auto-correlation and detailed balance, are not applicable 
for simulation-based inference methods like ABC.  This provides us the opportunity to massively parallelize ABC inference by following an embarrassingly parallel compute flow to form a batched version of ABC (see Fig.~\ref{f:abc_parallelized}), while still maintaining asymptotic convergence guarantees on the tolerance. Hence, we explicitly vectorize across parameter samples, while maintaining confidence that the true posterior will be approximated with accuracy equal to that of regular ABC. This vectorized simulation flow is well-supported by the TensorFlow framework~\cite{tensorflow2015-whitepaper}, allowing us to utilize the IPU's MIMD architecture, and the single-instruction-multiple-data (SIMD) architecture of GPUs.

For the epidemiology model (see Section~\ref{s:epi}), 
the original ABC algorithm would involve 
generating a single joint sample of parameters  $\theta$ of the size $[8]$, 
and generating data $D$ of size $[3, num\char`_days]$ 
(i.e. $A$, $R$, $D$ values for each of each of the days being simulated) 
and then compare with real data. In the new parallelized ABC algorithm, 
we sample multiple (batch size of $100$k or more) parameters $\theta$ 
by explicit vectorization, of size $[100000, 8]$,
simulate the resulting data $D_s$ which 
is also vectorized with size $[100000, 3, num\char`_days]$,
and compare the resulting simulated dataset with the ground truth dataset $D$
to a given tolerance level $\epsilon$.
At the end we calculate the number of accepted posterior samples and iterate
until a given required total number of accepted posterior samples is obtained.

% \centerline{\includegraphics[width=\linewidth]{Figures/abc_parallelize.PNG}}
% \caption{A conceptual representation of parallelizing ABC inference. (A)Overview of Vanilla ABC algorithm; (B) Parallel ABC - The sampling from prior, model simulation and distance computation are all embarrassingly parallelized, enabling support for hardware acceleration.
% }
% \label{f:abc_parallelized}
% \end{figure}

\subsection{Parallel Accept-Reject and Processing Accepted Samples}
\label{s:host}

One of the key considerations in the parallel ABC inference algorithm is the parallel accept-reject step of simulated samples. 
Further, we must consider the endpoint of the accepted samples. In some applications of ABC, the samples generated may be used directly on the accelerator. 
This might be true in Monte Carlo integration over the posterior, for example. 
However, in many cases we need to communicate the samples to the host for saving or further analysis. 
We are thus motivated to examine how efficient the process of parallel accept-reject and the communication of accepted posterior samples to host is on the different hardware platforms.

For efficient computation, the XLA backend had to be used. We observed a roughly 5x speedup with
it on the GPU. On the IPU, it is required by design. 
A caveat of the XLA approach is that 
the main algorithm needs
to return a fixed size of data.
For the given algorithm, we need to get the sampled parameters as well as their
tolerance values.
Collecting all data is not feasible.
For low tolerance values, we observed that the data would 
not fit into the working memory that comes with the GPU or IPU instance.
For higher tolerance values, we observed that the postprocessing 
can take even longer than the actual main processing.
Hence, different filtering strategies need to be applied to handle
the sample communication to host.

There are important distinctions between IPU and GPU in how accelerator-to-host transfer works. 
First, the IPU software includes \emph{outfeeds} into which data on the IPU can be enqueued and transferred to the host, 
all within an XLA-compiled graph. 
While the shapes of the tensors enqueued must be fixed for (XLA-)compilation, 
tensors do not need to be added to the outfeed at every run, i.e. for every batch of simulations. 
This means communication can be saved if no accepted samples were generated during the run. 
Further, we observe that for reasonable tolerances a very small fraction of samples are accepted, 
and therefore need to be transferred to the host. 
We thus reduce the overall communication volume, by breaking the batch of samples into chunks, 
and only transferring the chunk to the host if it contains at least one successful sample. For the experiments which follow, we use an chunk size of $10,000$ samples per IPU unless stated otherwise.

Conversely, data transferred from the GPU to host must be an output from the XLA graph on the GPU, 
either reducing the number of runs within each compiled graph, 
or requiring more samples to be stored on the accelerator before transfer.
We use the first approach, where we count the number of accepted samples each run. 
Independently, we store a fixed amount of samples per run with the lowest distances for that run (Top $k$). 
We then apply post-processing at the end of all the runs to accept the  posterior samples with distances less than the tolerance. The post-processing step is executed on the host.
This method introduces an additional design parameter $k$, 
which is the number of samples to return in runs that have non-zero acceptances. 
Based on the expected number of posterior samples per run, 
this parameter was set to $5$ for tolerance of $2E5$. 
For lower tolerances such as $5E4$, the parameter was set to $1$.
We did not use Top $k$ on the IPU because its outfeed 
feature allowed for a communication of all relevant samples, 
whereas the Top $k$ approach has a remaining low chance 
of removing relevant samples and introduces a hyperparameter that requires fine-tuning.

\underline{Ensuring Optimized Performance Across Platforms:}
other than how samples are returned, the parallelized ABC inference implementation is similar across both devices. In fact, we share exactly the same code for the simulation stages which we believe to be the bottleneck of the algorithm. We are therefore confident that our comparison between hardware platforms is fair. To ensure maximum efficiency in hardware, the compute-graph is compiled via XLA\footnote{See \url{https://www.tensorflow.org/xla}} for each device. For GPU, we use the XLA compiler provided in the TensorFlow framework. For IPU, we use its dedicated XLA compiler.
% developed by Graphcore. 
All experiments are run with float32 precision.

\section{Performance Analysis}
\label{s:performance}
% TODO mention max energy consumption index

In this section, we perform a detailed comparative analysis 
between the GPU and IPU implementations of ABC inference for the model described in Sections~\ref{s:epi} and~\ref{s:ABC}, and designed as per Section~\ref{sec:design}.
For this section, we used the case data for Italy\footnote{https://github.com/CSSEGISandData/COVID-19}. We performed simulation-based inference over $49$ days, where the task is  to accept a certain number of approximate posterior samples of the $8$ model parameters, with a given tolerance value. To obtain these posterior samples, the inference process has to be run for multiple times. As the tolerance is reduced, the the number of runs needed to obtain the same number of posterior samples increases. The number of samples simulated in every run of the inference process is referred to as the batch size. 

For this performance analysis we compare a C2 card containing 2 IPUs with the Tesla V100 GPU, given that both have the same TDP of 300W. We quantify key aspects of computation involved and explain how the differences relate to the unique architectures of the hardware platforms. 

For investigating performance, we use the TensorFlow Profiler\footnote{
See \url{https://www.tensorflow.org/tensorboard/tensorboard_profiling_keras}} 
to capture profiling data for the GPU. 
For the IPU, we use the PopVision Graph Analyzer Tool\footnote{
See \url{https://docs.graphcore.ai/projects/graphcore-popvision-user-guide/en/latest/}}.

\subsection{Runtime Comparison}
We first compare the runtimes 
between CPU (Xeon Gold 6248), GPU (Tesla V100), and IPU (Mk1).
We vary the acceptance threshold (tolerance) and the number of accepted posterior samples
and report the total time as well as the average time per run.
The latter is the more reliable metric because the total time 
can vary, largely due to the stochastic nature of how many posterior samples are accepted per run, causing stochasticity in how many runs are needed to generate a specified number of accepted samples which adds to the stochasticity in the total time.

Three things can be inferred from the results presented in Table~\ref{t:results}.
First, the IPU is consistently more efficient than GPU ($7.5\times$ speed up) and CPU ($30\times$ speed up)
over all three configurations.
Second, processing time scales linearly with the number of accepted samples for the values tested.
Last, decreasing the tolerance drastically increases the total processing time for all hardware configurations 
and could render GPU and CPU implementation infeasible in this regime,
especially when targeting multiple model iterations a day.

Due to the limited performance of the CPU, we focus the following analysis on GPU and IPU
to understand where this speedup comes from and how the algorithm behaves on these
two hardware platforms.

\begin{table*}[thbp!]

\begin{center}
\begin{tabular}{|c|c|c|c|r|r||r|r|r|}
\hline
\textbf{Device} & \textbf{Batch}& \textbf{Tole-}& \textbf{Accepted} &
\textbf{Total} & \textbf{Time per} & 
\multicolumn{3}{c|}{\textbf{Rel. \textit{Run} Perf. vs.}}\\
\textbf{} & \textbf{Size}& \textbf{rance}& \textbf{Samples} &
\textbf{Time (s)} & \textbf{Run (ms)} & \textbf{IPU} & \textbf{GPU} & \textbf{CPU}  \\
% \hhline{|=|=|=|=|=|=|=|=|=|}
\hline
\hline
2$\times$IPU & 2$\times$100k & 2E+05 & 100 & 
2.27 $\pm$ 0.29 & 4.71 $\pm$ 0.03 & 
1.00 & \textbf{7.47} & \textbf{30.87}\\
\hline
Tesla V100 & 500k & 2E+05 & 100 & 
14.87 $\pm$ 0.01 & 87.99 $\pm$ 0.04 & 
0.13 & 1.00 & 4.13\\
\hline
2$\times$ CPU & 1M & 2E+05 & 100 & 
67.87 $\pm$ 7.27 & 726.93 $\pm$ 6.01 & 
0.03 & 0.24 & 1.00\\
\hline
\hline
2$\times$IPU & 2$\times$100k & 2E+05 & \textbf{1000} & 
22.27 $\pm$ 0.76 & 4.72 $\pm$ 0.03 & 
1.00 & \textbf{7.24} & \textbf{29.56}\\
\hline
Tesla V100 & 500k & 2E+05 & \textbf{1000} & 
154.61 $\pm$ 0.04 & 85.47 $\pm$ 0.02 & 
0.14 & 1.00 & 4.08\\
\hline
2$\times$CPU & 1M & 2E+05 & \textbf{1000} & 660.48 $\pm$ 8.74 & 697.61 $\pm$ 4.87 & 
0.03 & 0.25 & 1.00\\
\hline
\hline
2$\times$IPU & 2$\times$100k & \textbf{1E+05} & 100 & 
77.07 $\pm$ 6.80 & 4.540 $\pm$ 0.001 & 
1.00 & \textbf{7.51} & \textbf{30.54}\\
\hline
Tesla V100 & 500k & \textbf{1E+05} & 100 & 
551.57 $\pm$ 0.20 & 85.29 $\pm$ 0.04 & 
0.13 & 1.00 & 4.32\\
\hline
% 2x CPU & 1M & \textbf{1E+05} & 100 & 
% 2383.10 $\pm$ 139.01 & 693.31 $\pm$ 4.71 & 
% 0.03 & 0.25 & 1.00\\
2$\times$CPU & 1M & \textbf{1E+05} & 100 & 
2383 $\pm$ 139 & 693.31 $\pm$ 4.71 & 
0.03 & 0.25 & 1.00\\
\hline
\end{tabular}
\caption{\textbf{Performance comparison:} The CPU is a Xeon Gold 6248. 
We provide mean and standard deviation from $10$ experiment repetitions. 
The relative run performance provides the speedup related to the time per run.}
\label{t:results}
\end{center}
\end{table*}

\begin{figure*}[htbp!]
\includegraphics[width=0.49\linewidth]{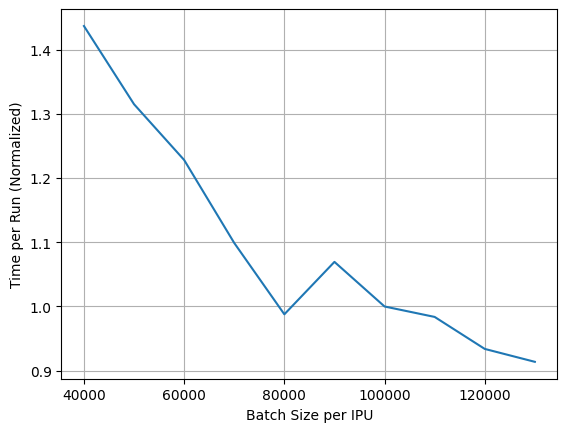}
\includegraphics[width=0.49\linewidth]{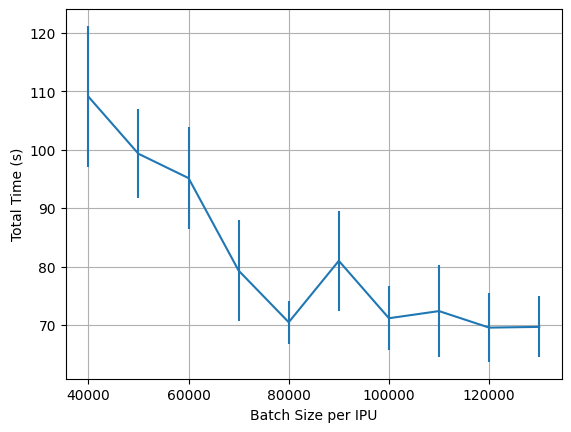}
\caption{Analysis of the influence of the batch size on the processing time of the IPU.
We generated $100$ accepted samples with a tolerance of $1E+05$ on $2$ Mk1 IPUs.
We normalized the time per run by dividing it by the batch size per IPU and multiplying it by $100.000$, such that the value at $100.000$ is $1.0$.
The standard error bars for the total time come from the $10$ repetitions.
}
\label{f:batch_size_ipu}
\end{figure*}

Table~\ref{tab:gpu_memory} for the V100 GPU and Table~\ref{tab:ipu_memory}
for the Mk1 IPU show different performance metrics 
depending on the chosen batch size for an evaluation with tolerance $2E5$
that collects $100$ samples.
For the IPU, the respective processing times are
also visualized in Figure~\ref{f:batch_size_ipu}.

As expected memory usage increases with increased batch size.
The GPU is capable of processing larger batch sizes than the IPU
due to having larger total memory.
The optimal performance is not obtained for a maximum batch size
but at $500000$ for the GPU and $120000$ for the IPU.

 \begin{table}[htbp!]
     \centering
     \begin{tabular}{rcccccc}
         Batch & Memory       & Active       & On-Chip          & Total & Time per\\
         Size & Used (MB/\%) &  Time ($\%$) & Resources ($\%$) & Time (s) & Run (ms)\\
         \hline
         $1\times10^5$ & 120 (0.82) &   55.5  & 90 & 15.85 & 19.88\\
         $2\times10^5$ & 240 (1.66) &   50.1  & 94 & 19.94 & 36.20\\
         $4\times10^5$ & 470 (3.25) &   52.2  & 97 & 16.55 & 69.53\\
         $5\times10^5$ & 590 (4.10) &   53.9  & 98 & \textbf{14.62} & 85.51\\
         $7\times10^5$ & 830 (5.70) &   54.4  & 98 & 15.59 & 118.10\\
         $10\times10^5$ & 1180 (8.10) & 55.3 & 99  & 15.11 & 167.89\\
     \end{tabular}
     \caption{The GPU performance profile for varying batch sizes at a tolerance of $2E5$ and $100$ collected samples.}
     \label{tab:gpu_memory}
 \end{table}

 \begin{table}[htbp!]
     \centering
     \begin{tabular}{rccccccc}
         Batch & Memory & Memory & Always
         & Active & Tile & Total & Time per\\
         Size & Used (MB) & Used ($\%$) & Live (MB) 
         & Time ($\%$) & Balance ($\%$) & Time (s) & Run (ms)\\
         \hline
         $2\times4\times10^4$ & 121 (160)  & 40 (52) & 31.5 & 83.5 & 97 & 3.13 & 2.67\\
         $2\times6\times10^4$ & 141 (160)  & 46 (52) & 28.4 & 84.6 & 96 & 2.62 & 3.44\\
         $2\times8\times10^4$ & 184 (187)  & 61 (62) & 33.4 & 85.4 & 96 & 2.26 & 3.71\\
         $2\times10\times10^4$ & 234 (234) & 77 (77) & 37.3 & 87.2 & 97 & 2.24 & 4.67\\
         $2\times12\times10^4$ & 265 (265) & 87 (87) & 38.8 & 87.9 & 98 & 2.10 & 5.28\\
         $2\times13\times10^4$ & 283 (284) & 93 (93) & 39.6 & 88.4 & 98 & 2.16 & 5.58\\
     \end{tabular}
     \caption{The performance profile for $2$ Mk1 IPUs for varying batch sizes at a tolerance of $2E5$ and with $100$ accepted samples. The value in the brackets denotes the memory use with ``gaps" -- memory on the tiles which cannot be used because it is reserved for different types of data than those in operation. ``Always Live" indicates how much memory must be accessible for all stages of algorithm, ``Active Time" is the overall percentage of tile cycles spent in execution and ``Tile Balance" measures how efficiently the operations are spread across tiles.}
     \label{tab:ipu_memory}
 \end{table}

\subsection{Communication Overhead and Host Processing}

% \subsection{Communications overhead (time spent in data transfers)}
To quantify the communication overhead due to transferring posterior samples from GPU to host, we ran the model twice -- both with and without sending samples from accelerator to host. We observed no discernible change in runtime due to our optimizations. Hence, data transfer overhead is not a significant factor in GPU acceleration. 
Consequently, the host-GPU interface bandwidth utilization was not investigated further.

% As discussed in Section~\ref{s:host}, the postprocessing of the data 
% from the IPU was placed onto the host.
% However, keeping in mind that small tolerance values are of interest for the application,
% this processing takes relatively little time.  % SN: should this paragraph be added to the section above? MMK: Some of the information here needs to go to the previous section, describing the implementations.

% TODO discuss processing overhead of outfeed data filtering on IPU in contrast to pure data sending. %SN: have started a bit of this above. 

As discussed in Section~\ref{s:host}, 
the IPU implementation transfers samples back to host in a fundamentally different manner. 
A batch of samples is enqueued into an output stream to host if any of the batch are accepted. 
For the IPU, we directly observe these enqueue ops and corresponding cycles in the execution traces.

Due to the execution of the enqueueing depending on at least one accepted sample in the batch, 
the fraction of cycles spent transferring to host depends on the tolerance. 
For drawing $100$ samples with a batch size of $100,000$ and a tolerance of $2E5$, 
we see these ops make up around $1.2\%$ of the overall number of cycles on the accelerator.   % SN: My calculation for 1.2%: StreamCopy outfeed enqueue takes up 133,000 cycles per run, but it is only run in ~ 200 of 400 runs (100 samples over 200 IPUs, mean 200 x enqueue ops). Total cycles for a run (with no outfeed enqueue is 5.75e6. Therefore, as a percentage of cycles: 133,000 * 200/400 / 5.75e6 = 1.2%
For $1E5$ tolerance, this decreases to $0.03\%$. % Similar calculation as for 1e5 tolerance. But here we do more runs for the same number of enqueue ops (accepted samples). Approximately 15000 runs, instead of 400. So fractional number of cycles is 133,000 * 200 / 15000 / 5.75e6
These observations corroborate with the reduction in wall clock time per run of 
$1.4\%$ when outfeed ops are removed, for a tolerance of $2E5$. %The value for $1E5$ is $0.56\%$.   %TODO (SN): find why the 1e5 tolerance values don't match up (run more repeats?)

%Portion of code running on CPU vs. on accelerator (GPU/IPU)}

For the best possible utilization of the accelerator platform, it is important to maximize the portion of code running on the accelerator and to minimize the portion running on host. For both accelerator setups, the samples generated on the accelerator are filtered on the host. We recorded the time spent running these postprocessing ops and present them in Table~\ref{t:host_postproc}.

It is clear that these host operations take a small fraction of the overall runtimes presented in Table~\ref{t:results} due to the optimization of the data transfer and not sending all data. Further, they take both a shorter wall-clock time and a smaller percentage of the overall runtime in the GPU configuration. This is likely to result from the difference in how many samples are sent to the host. In the GPU set up, only $5$ of most promising samples are sent for each run, whereas in the IPU case, a chunk of $10,000$ samples is sent if there is an accepted sample within it. This results in larger volumes of data being transferred to the host in the latter case, and therefore more host work to filter out the accepted samples. Comparing $100$ samples with $1000$ shows that the time required for postprocessing grows (approximately linearly) with the number of accepted samples.

\begin{table}
    \centering
    \begin{tabular}{|c|c|c|c|r|}
    \hline
    \textbf{Device} & \textbf{Batch Size}& \textbf{Tolerance}& \textbf{\# Accepted Samples} &
    \textbf{Postproc. Time (ms)}  \\
    % \hhline{|=|=|=|=|=|=|=|=|=|}
    \hline
    \hline
       Tesla V100  & 500k & 2E+05 & 100 & 15 (1.5\%)\\\hline
       2$\times$IPU & 2$\times$100k & 2E+05 & 100 & 70 (3\%)\\\hline
       2$\times$IPU & 2$\times$100k & 2E+05 & \textbf{1000} & 900 (4\%)\\\hline
       2$\times$IPU & 2$\times$100k & \textbf{1E+05} & 100 & 110 (0.1\%)\\\hline
    \end{tabular}
    \caption{Times spent postprocessing the data on the host, in milliseconds. The values in parentheses denote the percentage of total runtime spent on postprocessing.}
    \label{t:host_postproc}
\end{table}
% MMK: Seth please verify. We might also have to check what happens before the IPU gets running.
%SN: I get 2e5 100 samples: deq and postproc is 0.097s of 2.376s (4%). 1e5 1000 samples: 0.6s of  21.6s (2.8%).1e5 100 samples: 0.110s of 68.8s (0.16%).
%MMK: Which SDK did you use? Did you use 2 IPUs? %SN: sorry when I wrote "1e5 1000 samples" I meant "2e5 1000 samples". I'm using SDK 1.2.0-495c1aa368. Maybe the difference is due to CPU specs? I don't think the differences between our numbers change overall story, so maybe we should leave for now and come back if we have time? Maybe also we should do some repeats to test statistical significance? I've scripted this in run_cpu_vs_ipu.sh

In summary, the processing time on the host as well as
the times for data transfer to the host are small on both devices, cannot explain the performance difference in Table~\ref{t:results}.

%\subsection{Accelerator hardware resource utilization (GPU vs. IPU)}

\subsection{Memory Utilization}

 Table~\ref{tab:gpu_memory} displays details on memory usage
 for the GPU for different batch sizes.
 For a batch size of $500K$, the memory utilization of the GPU is $0.59$GB/$14.42$GB ($4.1\%$) and results in the best average runtime.
 While the GPU's large off-chip memory enables large batch sizes, the runtime does not benefit for sizes larger than $500K$. This may be due to the GPU's memory hierarchy. The parameter array, which is of size $[500000, 8]$ is around $15$MB at single precision, which is close to the total L1+L2 cache of $16$MB. Hence any batch size greater than $500$k would result in a parameter array larger than the combined L1 and L2 caches, needing frequent main memory look-ups and some serialisation of the computation, and hence not providing any additional benefit with increasing batch size.
 Keeping in mind that the aggregated data covers $500000$ simulations,
 of $49$ days each with $6$ population variables,
 the GPU is not able to hold all the simulation data 
 in the L1 and L2 caches and 
 has to obtain at least part of it from the main memory.\footnote{
 The data array consists of $500.000*49*6=147.000.000$
 parameters which in the $32$ bit setup requires around
 $560$ MB of memory and exceeds the L1+L2 cache size.}

 % How much are the caches used?
 % How much communication to memory occurs?
 % How efficient is the memory handling?
 
\begin{figure}[htbp!]
     \centering
     \includegraphics[
     width=\linewidth, trim={0 0 0 1.5cm},clip]{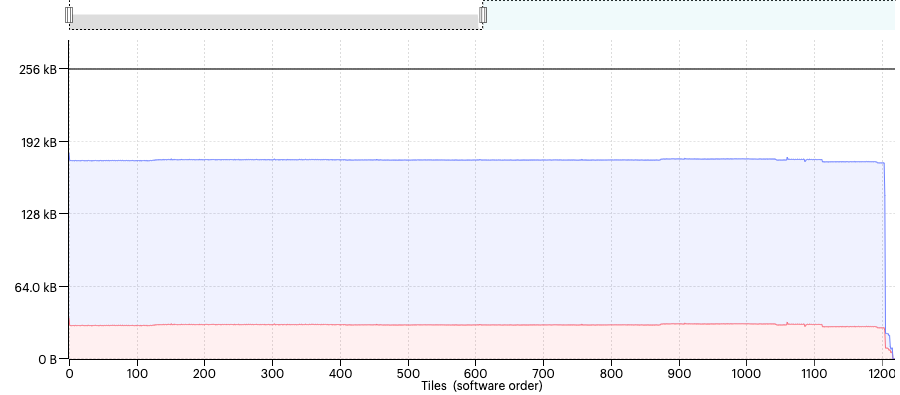}
     \caption{The (maximum) memory consumption (vertical axis) 
     across the $1216$ tiles (horizontal axis) of a MK1 IPU for a batch size of $100,000$. 
     The horizontal line denotes the maximum available tile memory.
     The pink area represents always live memory, 
     whereas the blue indicates maximum transient memory use on a tile level.}
     \label{fig:ipu_tile_mem}
 \end{figure}

 The memory utilisation of IPUs varies with the batch size as shown in Table~\ref{tab:ipu_memory}. 
 There is clearly an increase in memory consumption with increasing batch size. 
 However, the two are not proportional, 
 indicating that significant memory is consumed by quantities 
 other than the arrays of samples generated, 
 whose size is less dependent on the batch size. 
 Further investigation found that some of this additional memory is consumed 
 by code residing on tiles required to describe the local computation. 
 The distribution of memory over tiles (Fig.~\ref{fig:ipu_tile_mem}) is close to uniform, 
 suggesting efficient memory use in general and effective load balance between tiles on the IPU.  
 We do not observe any variation in memory consumption with tolerance. This is reasonable since the tolerance as well as the number of accepted/requested samples does not change the computational graph.
 
\begin{figure}[htbp!]
     \centering
     \includegraphics[
     width=\textwidth, trim={0 0 0 1.5cm},clip]{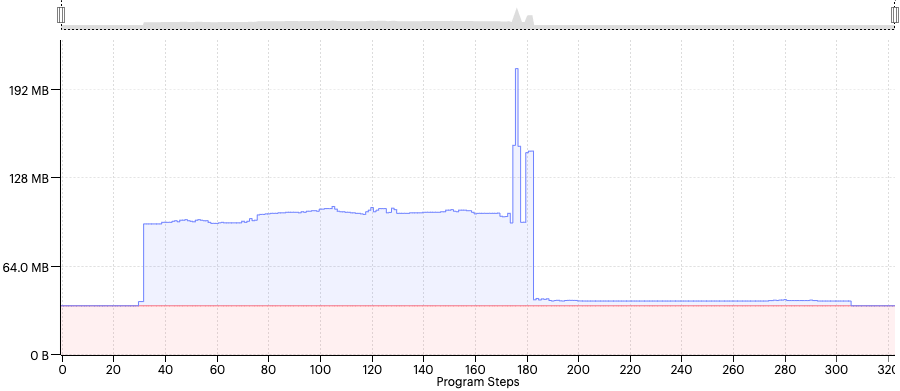}
     \caption{The memory liveness for a MK1 IPU at a batch size of $100,000$, tolerance 2E05. The horizontal axis is the program step, vertical is live memory. The pink area represents always live memory, whereas the blue indicates transient memory use.}
     \label{fig:ipu_live}
 \end{figure}

 We can explore memory consumption on the IPU further by looking into memory \textit{liveness}: 
 the memory consumption at different algorithmic steps, and the extent to which memory is constantly in use or only allocated temporarily. 
 The memory liveness of the IPU implementation
 of one single simulation run
 is illustrated in Fig.~\ref{fig:ipu_live} (batch size $100,000$, tolerance $2E05$). 
 The plot indicates that there is a significant amount of memory always allocated (pink area, ``always live''), 
 but the peak memory liveness is around six times greater than the always live amount. 
 The most prominent peaks were found to be caused by calculating the Euclidean distance between each of the $100,000$ samples and the real data.
 In unpublished results, we experimented with
 breaking down the distance calculation 
 by incrementally adding differences up for each day.
 This decreased performance.

 We observe that the IPU is making much better use of its memory with $77\%$ usage.
 It is able to keep instructions and processed data in memory.
 In contrast, the GPU is not able to hold code and data in cache for the computations which leads to a large overhead in loading those from main memory.
 Also, the large memory of the GPU does not pay off in this application
and only $4\%$ is used in the best-performing configuration.

\subsection{Processing Load Distribution}

Table~\ref{tab:ipu_memory} shows 
that for a batch size of $100000$, $97\%$ of the on-chip resources are used.
Interestingly, the GPU shows a similar coverage of $98\%$ in Table~\ref{tab:gpu_memory}.
Note however, that at least for single precision, the two IPUs come with $62$ TFLOPS,
in contrast to the $14$ TFLOPS of the GPU.

Also, the table comparison shows that the GPU active time is
$53.9\%$ for the best batch size in contrast to $87\%$ for the IPU.
The idle time is likely due to time spent waiting for loading code for kernel launch.
For the IPU, we observed in detailed profiles that most operations were divided over most of the tiles and run in parallel. However a very small subset of operations
were processed on a fraction of the tiles while the others waited.
The remaining $13\%$ of the cycles were used for the synchronization.
Any other processing like the actual data exchange were marginal.
The large waiting time from the GPU might come from it having to load
the code and data from the main memory and not the cache.

The $87\%$ of active time by the IPU can be further broken
down on a compute set level as shown in Table~\ref{tab:ipu_cycles}.
It is striking that even on the tile level,
there is a significant number of operations that rearrange the data (PreArrange, OnTileCopy, slice, update, PostArrange, Transpose, OnTileCopyPre).
Altogether they constitute $50\%$ of the cycles.
So here, the IPU can also benefit from the high
memory bandwidth within a single tile.

For the GPU, the breakdown can be seen in Table~\ref{tab:gpu_kernels}.
Due to XLA optimization, single operations cannot be
inferred anymore but it seems that most 
are fused into one single operation that uses $72.3\%$
of the processing time.

We may interpret the difference in runtime scaling with batch size in a number of ways (comparing Tables~\ref{tab:gpu_memory} and~\ref{tab:ipu_memory}). First, the fact that the GPU runtime does not benefit from larger batch sizes may be due to being bottlenecked by memory bandwidth in the cache hierarchy where the IPU has high bandwidth, local memory. Second, we observe that the two IPUs provide around $4\times$ the number of FLOPS of the GPU. Thus the GPU processing capacity maybe exhausted at the smallest batch size tested, resulting in a near linear increase in time per run with increasing batch size. We believe a combination of these factors contribute to the difference in scaling and the speedup of IPU over GPU.
% Putting together the increased Flops and more efficient memory usage of the MK1 IPU
% makes the performance advantage of $7.5x$ reasonable.

 \begin{table}[htbp!]
     \centering
     \begin{tabular}{lr}
         Compute Set Suffix & Cycles ($\%$)\\
         \hline
            Power &	24.3\\
            PreArrange &	22.5\\
            Add &	10.8 \\
            OnTileCopy &	10.1\\
            slice &	9.5\\
            Multiply &	4.1\\
            update &	4.0\\
            Clamp &	2.3\\
            Sqrt &	1.9\\
            PostArrange &	1.8\\
            Transpose &	1.5\\
            Reduce &	1.4\\
            normal &	1.4\\
            Convolve &	1.2\\
            Floor &	1.0\\
            OnTileCopyPre &	0.7\\
            Divide &	0.7\\
            Others & 0.9
     \end{tabular}
     \caption{The distribution of non-idle IPU cycles for $100$ accepted samples and $2E5$ tolerance.}
     \label{tab:ipu_cycles}
 \end{table}

 \begin{table}[htbp!]
     \centering
     \begin{tabular}{lr}
         XLA Kernel & Runtime ($\%$)\\
         \hline
            fusion\textunderscore5 &	72.3\\
            fusion\textunderscore9 &	8.9\\
            volta\textunderscore sgemm &	6.1 \\
            fusion\textunderscore8 &	4.8\\
            fusion\textunderscore5\textunderscore1 &	2.6\\
            fusion\textunderscore10 &	1.6\\
            fusion\textunderscore11 &	1.4\\
            fusion\textunderscore64 &	1.2\\
            fusion\textunderscore60 &	0.6\\
            broadcast\textunderscore682 &	0.4\\
     \end{tabular}
     \caption{The distribution of non-idle XLA kernels for $100$ accepted samples and $2E5$ tolerance for GPU.}
     \label{tab:gpu_kernels}
 \end{table}

% \subsection{XLA comparison (removal candidate)}

% If possible generate computational graphs compiled for GPU and IPU by XLA, compare them

% can this be done?

% % MMK: on the IPU, I would use:

% % os.environ["XLA_FLAGS"] = "--xla_dump_to=tmp_xla_{} ".format(
% %     np.random.randint(2, 101))
% % os.environ["XLA_FLAGS"] += " --xla_dump_hlo_pass_re=forward-allocation "
% % os.environ["XLA_FLAGS"] += " --xla_hlo_graph_sharding_color "
% % os.environ["XLA_FLAGS"] += " --xla_dump_hlo_as_text "

% % MMK: Maybe that also works on GPU? However, I have doubts that the graphs will be really
% % useful and it might be impossible to compare. I was thinking the same

% % MMK:  I compared the files and there is no use. The code that goes to the devices is significantly different
%vs. speedup comparison between GPU and IPU}

\subsection{Scalability of Model to Multiple IPU Devices}

\begin{figure}[htbp]
\centerline{\includegraphics[width=0.6\linewidth]{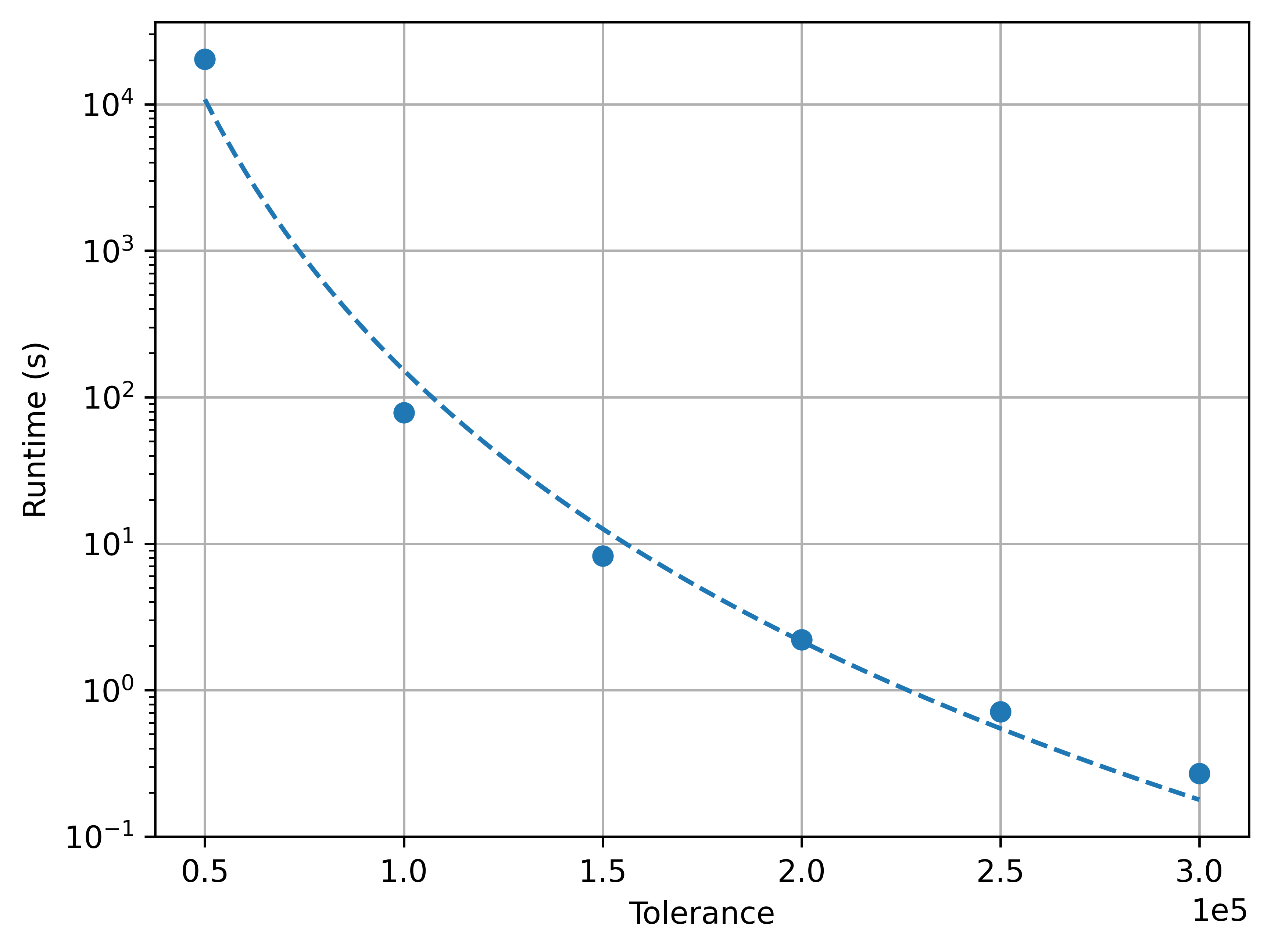}}
\caption{Computation time dependent on tolerance
for aggregating $100$ samples on two Mk1 IPUs.
Note the exponential scaling on the 
horizontal axis and the
super-exponential increase of processing time with decreasing tolerance.
}
\label{f:tolerance_plot}
\end{figure}

For each dataset, a different tolerance is appropriate, 
because it depends on noise in the data, 
predictability of the progress from the epidemiological model,
and the number of days.
For the Italian dataset, high quality samples can be obtained using a tolerance of $5E+04$.
For this tolerance however, obtaining the good parameters estimates takes too much time for
experimentation.
This can be seen in Figure~\ref{f:tolerance_plot}.
On two Mk1 IPUs, the processing takes more than $5$ hours ($18000$s).
Note that a representative sample of $1000$ samples
would consequently take more than two days.
Despite our acceleration, this is considered prohibitive during development.

\begin{table*}[thbp!]
\caption{\textbf{Scalability Analysis:} 
For the speed up calculation the time per run is used,
corrected by the ratio of the batch sizes.}

\begin{center}
\begin{tabular}{|c|c|c|c|c|r|r||r|r|r|}
\hline
\textbf{Device} & \textbf{Batch}& \textbf{Tole-}& \textbf{Accepted} &
\textbf{Chunk} &
\textbf{Total} & \textbf{Time per} & 
\multicolumn{3}{c|}{\textbf{Speed up}}\\
\textbf{} & \textbf{Size}& \textbf{rance}& \textbf{Samples} & \textbf{Size} &
\textbf{Time (s)} & \textbf{Run (ms)} & \multicolumn{3}{r|}{rel. to 2xIPUs} \\
\hline
\hline
% TODO: Split into separate table in separate section
\hline
% 2xIPUs & 2x100k & \textbf{5E+04} & 100 & 
% 19331 $\pm$ 1142 & 4.52 $\pm$ 0.00 & 
% \multicolumn{3}{c|}{}\\
2xIPUs & 2x100k & {5E+04} & 100 & 2x10k &
20354 & 4.53 & 
\multicolumn{3}{c|}{}\\
\hline
4xIPUs & 4x100k & {5E+04} & 100 & 4x10k &
9056 & 4.6 & 
\multicolumn{3}{r|}{1.97}\\
\hline
8xIPUs & 8x100k & {5E+04} & 100 & 8x10k &
4536 & 4.7 & 
\multicolumn{3}{r|}{3.85}\\
\hline
% 16xIPUs & 16x100k & \textbf{5E+04} & 100 & 
% 2422 $\pm$ 150 & 4.57 $\pm$ 0.00 & 
% \multicolumn{3}{c|}{7.9x faster than 2xIPUs}\\
16xIPUs & 16x100k & {5E+04} & 100 & 16x10k &
2355 & 4.91 & 
\multicolumn{3}{r|}{7.38}\\
\hline
\hline
8xIPUs & 8x100k & {5E+04} & 100 & 8x100k &
4091 & 4.49 & 
\multicolumn{3}{r|}{4.04}\\
\hline
16xIPUs & 16x100k & {5E+04} & 100 & 16x100k &
2189 & 4.52 & 
\multicolumn{3}{r|}{8}\\
\hline
\end{tabular}
\label{t:scaling}
\end{center}
\end{table*}
To overcome this limitation of processing time, 
we test how computation can be accelerated by distributing it across multiple devices.
Note that sample generation could run embarrassingly parallel and scale perfectly.
In the following experiment however, we want to determine 
how much the communication between devices slows down the processing and how fast we can get results.
One relevant parameter here is the chunk size parameter.
Before data is sent from IPU to host, 
it is divided in respectively sized data chunks
and only relevant chunks are sent to host.

The results are provided in Table~\ref{t:scaling}.
There are three key findings.
First, with $16$ IPUs, we got the result in less than $40$ minutes.
This is fast enough for having multiple iterations and experimenting
with the model.
Second, the performance scales linearly with the number of IPUs.
Third, performance scales better 
when we are not chunking the data, 
i.e., chunk size equals to batch size.
Chunking comes with additional synchronization between the IPUs
but reduces the time for the postprocessing.
If there is no chunking, the final dataset that gets filtered on the 
host occupied more than $10\%$ of the host memory,
whereas the chunked data occupied significantly less data.
The respective memory acquisition takes between 
$0.5-1$ second for $100$ samples.
With high tolerances, we have short processing times such
that performance benefits from the chunking.
However with small tolerances, the chunking requires more processing
and it is more efficient to do the processing on the host.

% TODO: Transition to next section by pointing out practical relevance of fast compuations

\section{Epidemiological Model Analysis}
\label{s:analysis}

\begin{figure*}[htbp!]
\begin{tabular}{ccc}
 \footnotesize Italy & 
 \footnotesize New Zealand & 
 \footnotesize USA\\
 \includegraphics[width=0.3\linewidth]{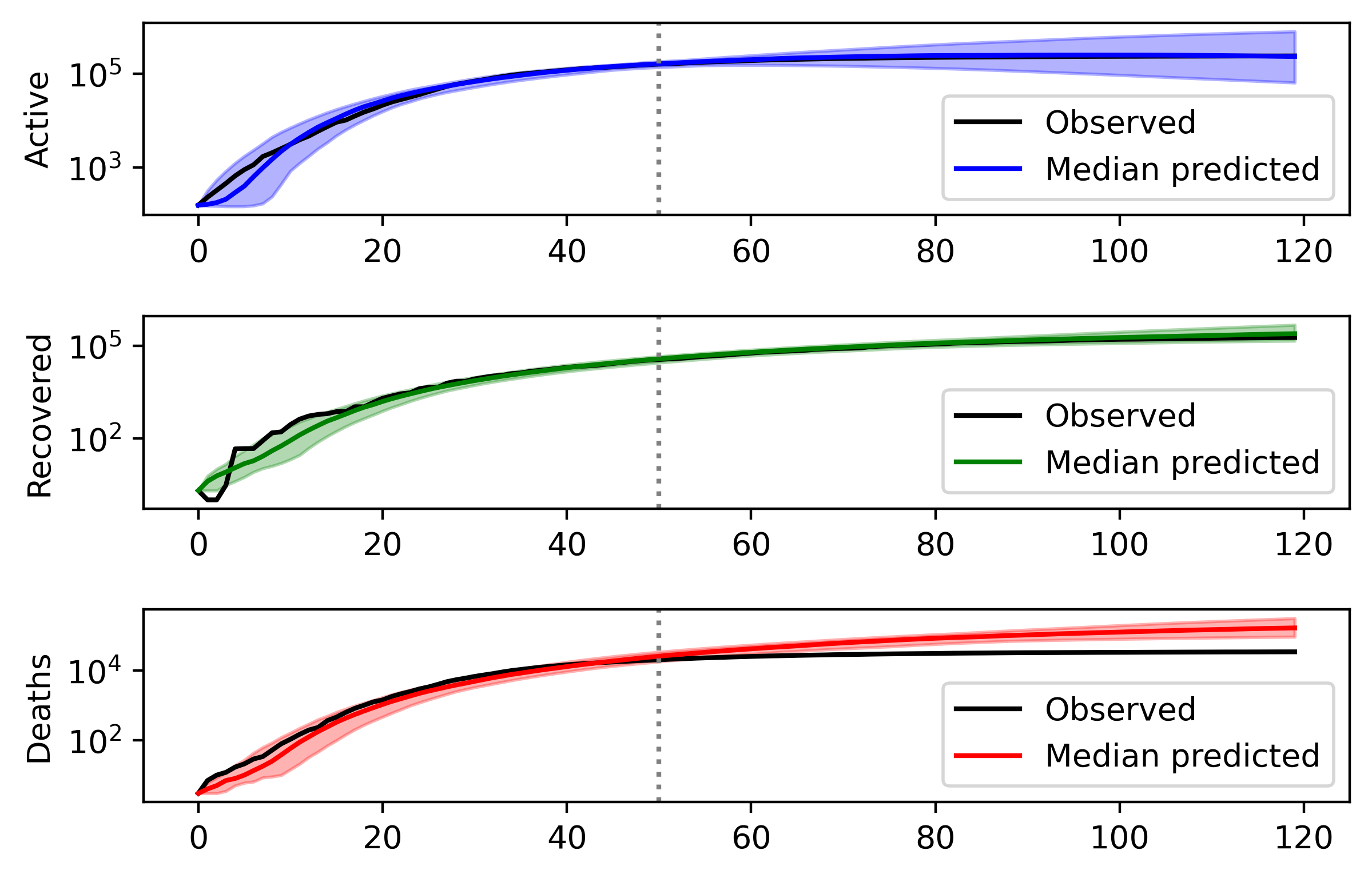} &
 \includegraphics[width=0.3\linewidth]{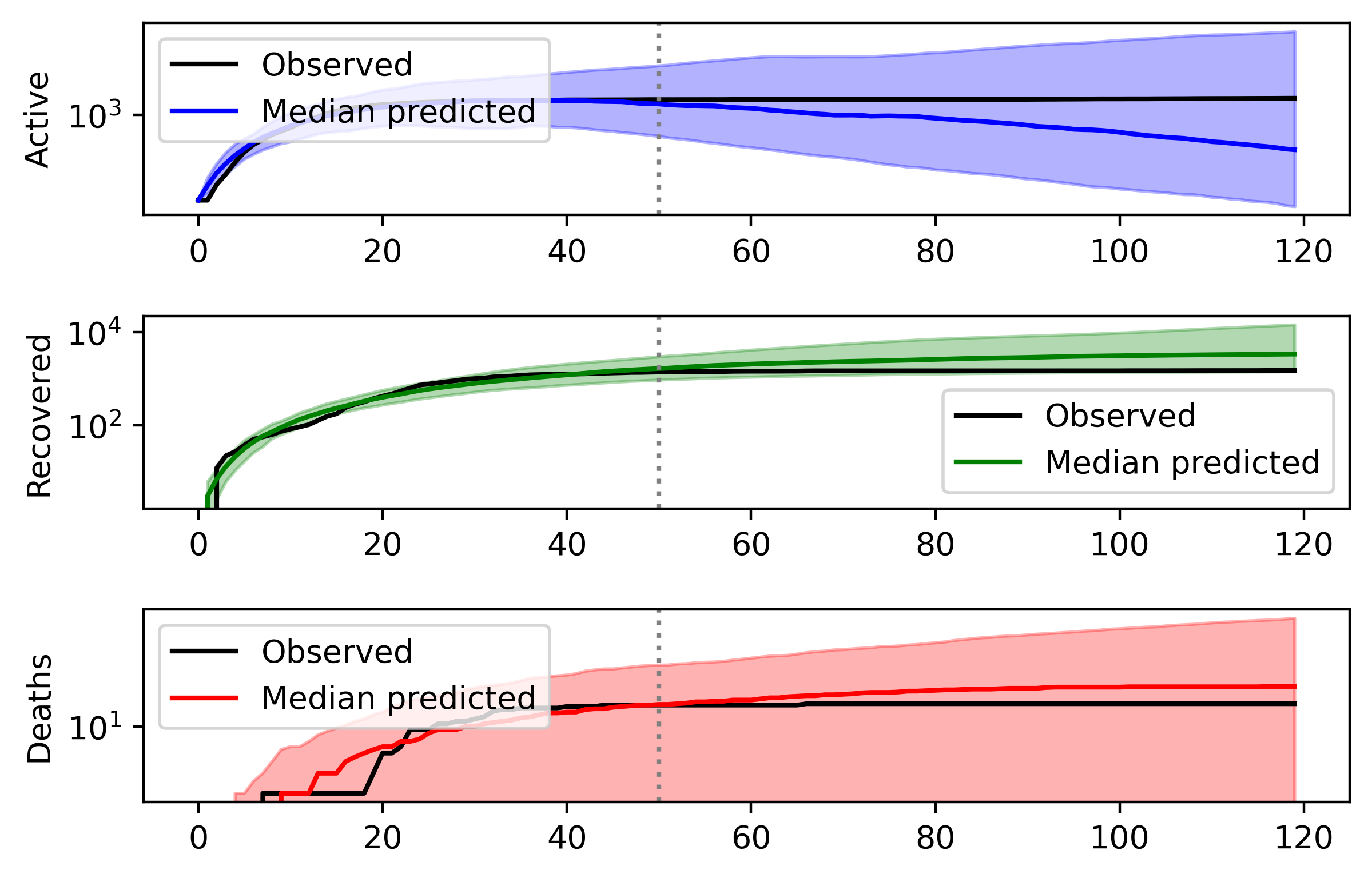} &
 \includegraphics[width=0.3\linewidth]{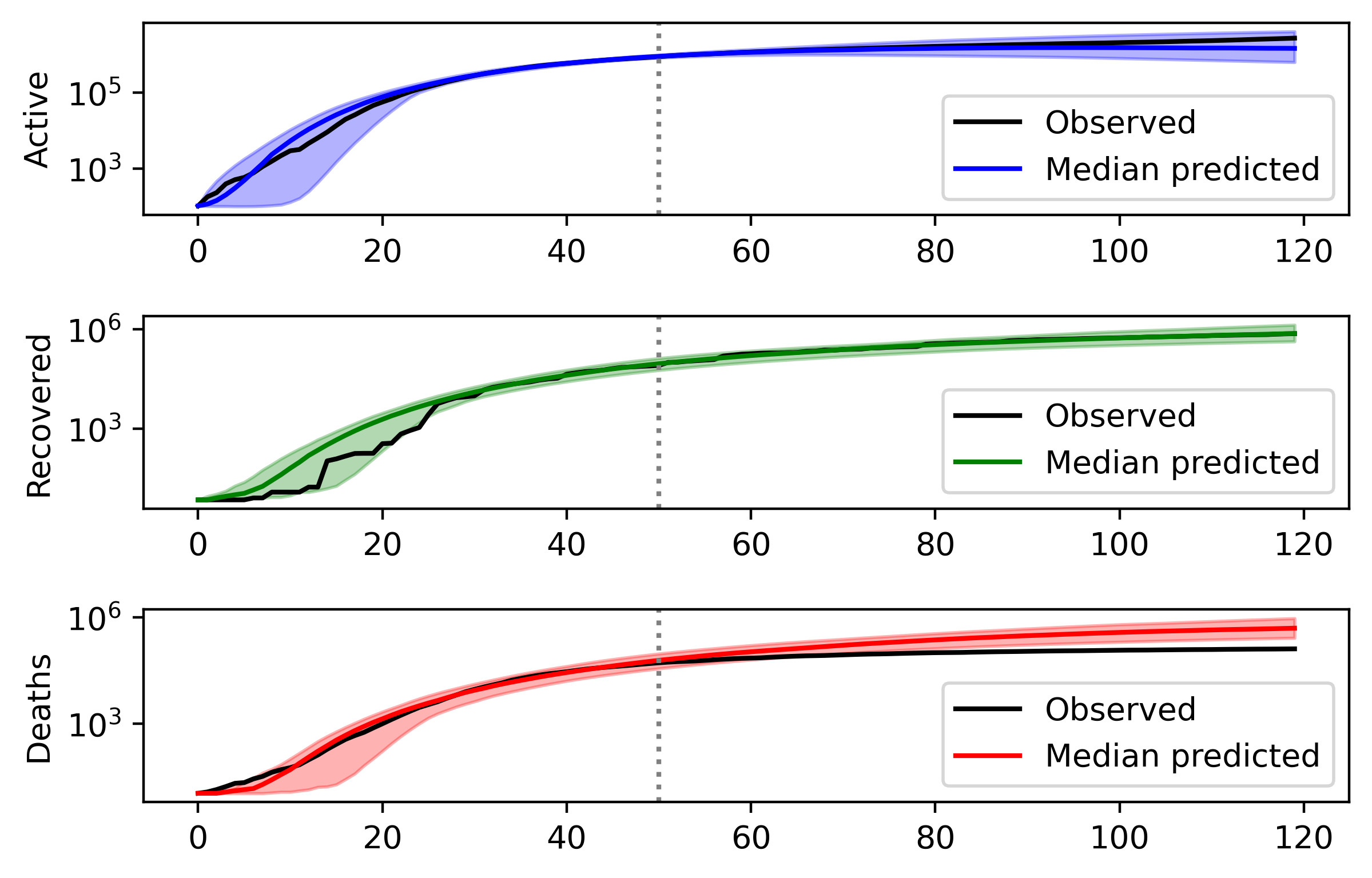}
\end{tabular}
\caption{Projected 120-day trajectories of Active, Recovered, and Death cases for Italy, New Zealand, USA. Generated by simulating the model with $100$ accepted samples. These samples were obtained by training the model with data for $49$ days.
The color-shaded area covers the uncertainty from the different projected trajectories
between $5^{th}$ and $95^{th}$ percentile.
The x-axes show the number of days.
}
\label{f:trajectories}
\end{figure*}

In this section, we present an example of our approach working in practice, take a look at the actual results and
show how they can be used to compare different countries.
For our analysis we perform parallel ABC inference to fit the epidemiology model to data for $49$ days (after the first day with $100$ detected cases), across three countries (Italy, New Zealand, USA). These countries were chosen to capture a diverse set of the pandemic outcomes and population sizes. The data was obtained from the Johns Hopkins dataset\footnote{https://github.com/CSSEGISandData/COVID-19}. To fit the model to data, the algorithm was run until we accepted 100 samples from the approximate posterior. 
Then, we use the obtained posterior samples to perform $100$ simulations, generating predictions for $120$ days (after the onset).
Note that it is not appropriate to choose the same tolerance for each country.
Na\"{i}vely scaling the tolerance by the population is also not advised.
Each country has a different level of noise in the data, so
the simulation model might better fit to policies in some countries,
in addition the virus has different characteristics in each country.
Thus the tolerance had to be adjusted on an individual basis.
We used $16$ Mk1 IPUs with no chunking, and a batch size of $100k$ per IPU. The predictions, with tuned tolerances, are plotted along with the true data in Figure~\ref{f:trajectories}. 

We can see that the model fits the data well and generates reasonably accurate forecasts.
The error margins for New Zealand probably appear large relative to the
small case numbers. Reducing the tolerance from original $2000$ to $1250$ 
did not significantly change this. We note also that our Gaussian approximation to the Poisson sampling may be not satisfactory for such low case numbers.

The respective performance results for obtaining the 
parameters are summarized in Table~\ref{t:trajectories}, as 
well as the means of the approximate posterior parameter samples.

\begin{figure*}[htbp!]
\begin{tabular}{ccc}
 \footnotesize Italy & 
 \footnotesize New Zealand & 
 \footnotesize USA\\
\includegraphics[width=0.3\textwidth,height=0.9\textheight,keepaspectratio]{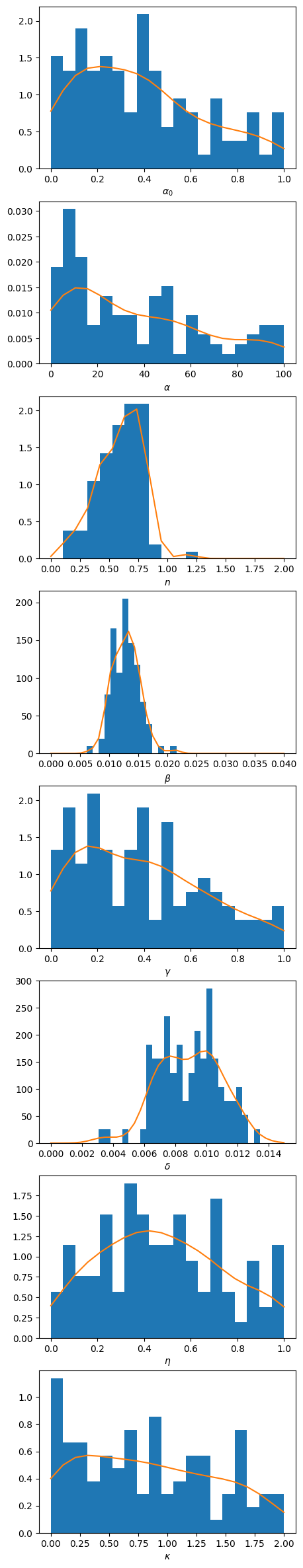} &
\includegraphics[width=0.3\textwidth,height=0.9\textheight,keepaspectratio]{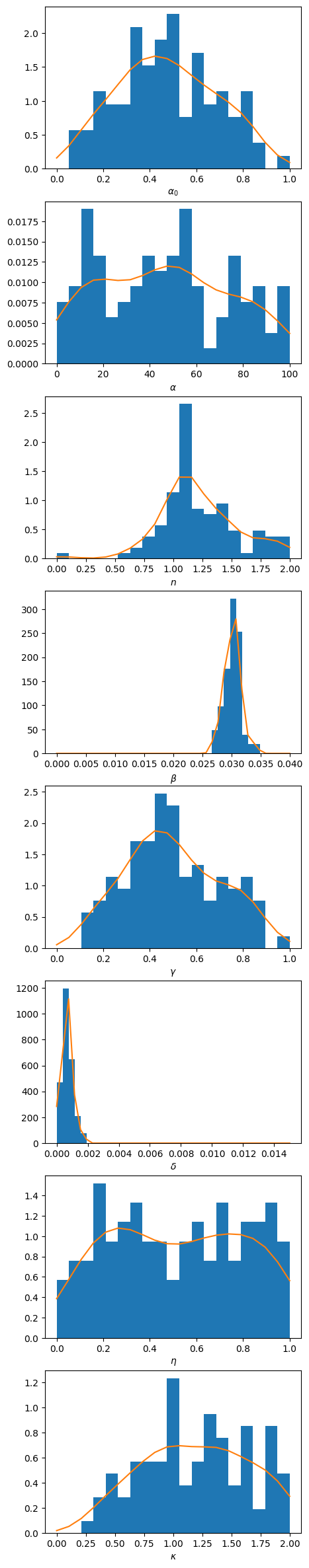} &
\includegraphics[width=0.3\textwidth,height=0.9\textheight,keepaspectratio]{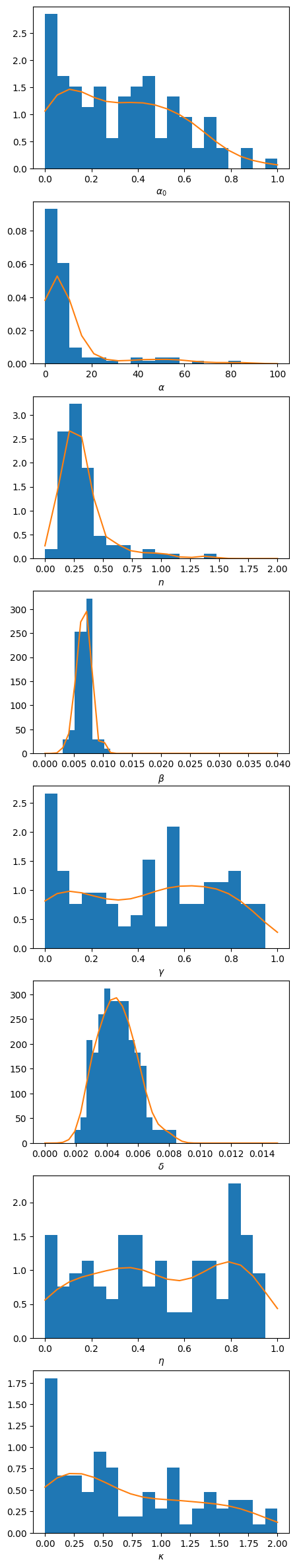}
\end{tabular}
\caption{Histograms for $100$ accepted samples of the $8$ model parameters for Italy, New Zealand, and USA.
}
\label{f:histograms100}
\end{figure*}

\begin{figure*}[htbp!]
\begin{tabular}{ccc}
 \footnotesize Italy & 
 \footnotesize New Zealand & 
 \footnotesize USA\\
\includegraphics[width=0.3\textwidth,height=0.9\textheight,keepaspectratio]{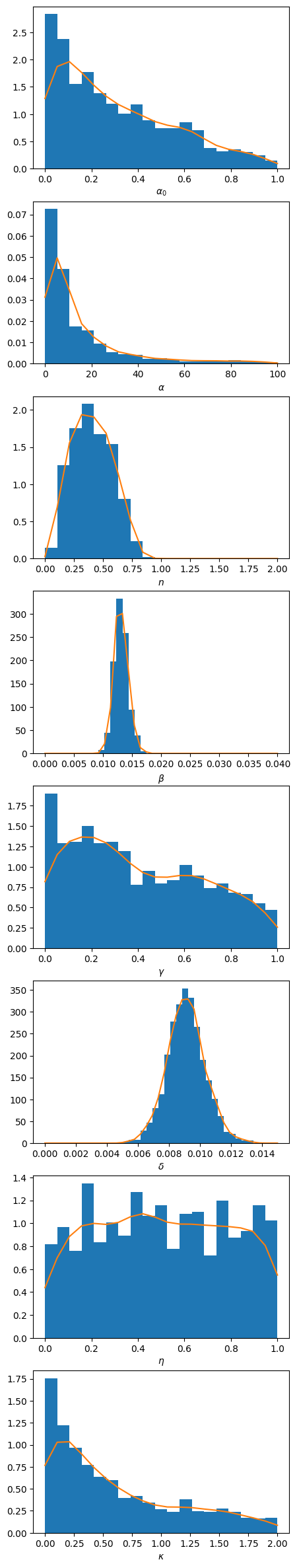} &
\includegraphics[width=0.3\textwidth,height=0.9\textheight,keepaspectratio]{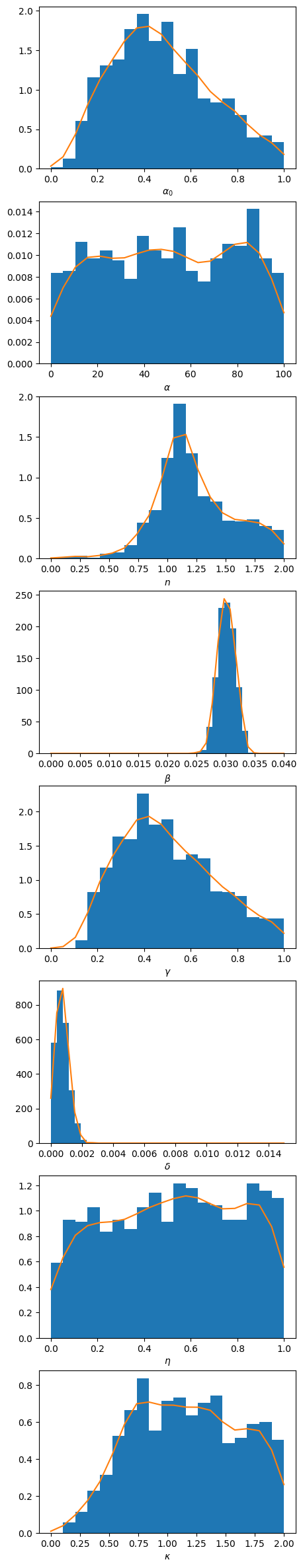} &
\includegraphics[width=0.3\textwidth,height=0.9\textheight,keepaspectratio]{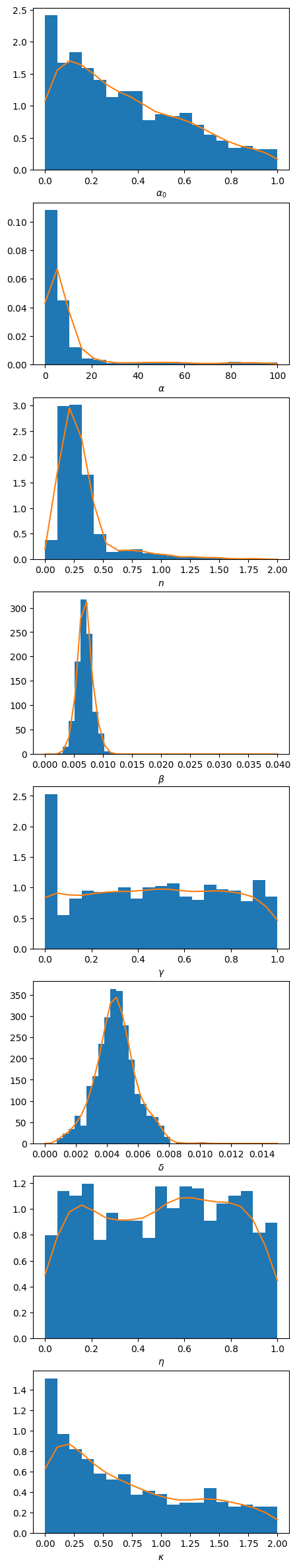}
\end{tabular}
\caption{Histograms for $1000$ accepted samples of the $8$ model parameters for Italy, New Zealand, and USA.
}
\label{f:histograms1000}
\end{figure*}

It is striking that the average recovery rate in New Zealand
is twice as high as the recovery rate in Italy which, in turn, is
twice as high as in the USA.
The average fatality rate of Italy is almost twice as
high as in the USA and almost ten times higher 
than in New Zealand.
The exponent $n$ for the infection rate in New Zealand is
more than twice that in the other countries.
These three signs indicate that the pandemic was well controlled and the cases well treated in New Zealand, relative to US or Italy.

Figure~\ref{f:histograms100} and Figure~\ref{f:histograms1000}
show the histograms of $100$ and $1000$ posterior samples respectively.
They show that at least some of the posterior parameter distributions are not Gaussian
and that the aforementioned differences in the parameter averages
are actually significant.
Also, we can see that the histograms for $1000$ accepted samples
are much less noisy than the ones for $100$ accepted samples.
For example for the model parameters $\beta$ and $\delta$ in case of Italy, with $100$ samples it is unclear weather the posterior is uni-modal or bi-modal, as the approximate posterior is consistent with both. With $1000$ samples, it becomes clear that they are indeed uni-modal. 
% For the parameter alpha, the distribution for 
% Italy and New Zealand is quite unclear with 100 samples
% but becomes clear with 1000 samples.
Considering the parameter $\alpha$ for Italy and New Zealand, with 100 accepted samples, the posteriors seem identical, but with 1000 accepted samples, the differences among them are quite clear. 

\begin{table}[htbp!]
     \centering
     \begin{tabular}{rccccccccccc}
         & Tole- & Run & Accepted & \multicolumn{8}{c}{Average}\\
         Country     & rance & time (s) & 
         samples & $\alpha_0$ & $\alpha$ & $n$ & $\beta$ & 
         $\gamma$ & $\delta$ & $\eta$ & $\kappa$ \\
         \hline
         Italy       & 5e4       & 2107 & 100 &
         0.384 & 36.054 & 0.595 & 0.013 & 0.385 & 0.009 & 0.477 & 0.830\\
         New Zealand & 1250      & 1330 & 100 &
         0.474 & 46.603 & 1.223 & 0.030 & 0.499 & 0.001 & 0.520 & 1.198\\
         USA         & 2e5       & 4420 & 100 & 
         0.329 & 10.667 & 0.322 & 0.007 & 0.435 & 0.005 & 0.490 & 0.716\\
         \hline
         \hline
         Italy       & 5e4       & 22954 & 1000 &
         0.309 & 15.012 & 0.406 & 0.013 & 0.407 & 0.009 & 0.507 & 0.603\\
         New Zealand & 1250      & 13846 & 1000 &
         0.480 & 50.932 & 1.238 & 0.030 & 0.504 & 0.001 & 0.527 & 1.172\\
         USA         & 2e5       & 46339 & 1000 & 
         0.342 & 11.051 & 0.319 & 0.007 & 0.468 & 0.005 & 0.501 & 0.706\\
           \\
     \end{tabular}
     \caption{Parameter averages for different countries with total infection rate 
     $g_{(A,R,D)} = \alpha_0 + \frac{\alpha}{1 + (A+R+D)^n}$,
     recovery rate $\beta$, positive test rate $\gamma$,
     fatality rate $\delta$,
     testing protocol effectiveness $\eta$, and
     initial unobserved cases rate $\kappa$.
     }
     % base infection rate, infection rate change, infection rate change exponent, 
     \label{t:trajectories}
 \end{table}

\newpage

\section{Discussion and Conclusion}
\label{s:conc}

This work demonstrates the potential of hardware-accelerated simulation-based inference algorithms for stochastic epidemiology models. We implement parallelized ABC inference on a Xeon CPU, Tesla V100 GPU, and MK1 IPU. Compared to CPU, the proposed parallel ABC version is $\approx4\times$ faster on GPU, and $\approx30\times$ faster on 2xIPUs. We performed extensive analysis to quantify the performance difference across GPU and IPU implementations. Our analysis indicates a three-fold advantage of IPU over the GPU: 

i) The GPU runtimes do not benefit from batch sizes above 500k, despite this being only $\approx 8\%$ utilization of the main memory. Our calculations suggest the on-chip caches of the GPU ($6$MB L2 + $10$MB L1)  are too small to hold the data needed for simulation, and therefore constant interaction with main memory is a probable bottleneck. Conversely, the IPU has large on-chip memory ($300$MB) and we see that increasing batch size consistently provides better performance, up to $90 \%$ memory utilization.
%i) The IPUs have much larger on-chip "cache" memory ($300$MB) vs. GPU ($6$MB L2 + $10$MB L1). Hence, while GPUs could potentially fit much larger batch sizes into their main memory ($14$GB), increasing batch size shows no benefit already at $\approx 8\%$ of memory utilization. On the other hand, increasing batch size in IPU consistently provides better performance, up to $90 \%$ memory utilization.

% ii) The IPU architecture is MIMD, while GPU architecture is SIMD. For the application considered, the MIMD architecture was able to distribute computation across on-chip resources on the IPU much better than the SIMD architecture in GPU. The overhead of deploying code to GPU was $\approx43\%$.

ii) The GPU is fetching the instructions and the data 
from memory to the accelerator chip and back.
The overhead of deploying code to GPU was $\approx43\%$.
In contrast, data and instructions reside in the IPU memory 
and no transfer of instruction code is required ($0\%$ overhead).

iii) The IPU has a much higher within-device memory bandwidth vs. the GPU ($45$ TB/s  vs. $900$ GB/s). The IPU also has much greater compute throughput ($62$ TFLOPS for 2xIPU vs. $14$ TFLOPS for GPU).
These benefits lead to much faster computations in IPU  vs. GPU.  

Hence, we believe the higher bandwidth, closeness of memory to compute, 
and better compute power allow 2xIPU to be $\approx7\times$ faster than Tesla V100 GPU. 

To assess the scalability of the proposed approach 
to multiple devices, 
we demonstrate parallel ABC inference 
over the epidemiology model from 2x through 16x IPUs. 
We observe that the scaling overhead is between 
$0\%$ (no on-device sample post-processing) 
to $8\%$ (with on-device sample post-processing) 
depending on the configuration. 
The increase in post-processing on-device saves time in host processing, so this provides with a trade-off between efficient scaling vs. reducing host processing times.

We also perform analysis on the epidemiology model across three countries (Italy, USA and New Zealand). We obtain accurate predictions for each which are plotted. We also observe the inferred parameters for the three countries. In general, the model tracks actual data well. To perform this analysis, the total runtime using a $16$-IPU system was $\approx 1.9$ Hours. We estimate that doing a similar analysis over $16$ Xeon CPUs would have taken $\approx 57$ Hours, while $8$ V100 GPUs would have taken $\approx 15$ hours.

We believe that the parallelized ABC inference proposed in this work would generalize well across a wide variety of epidemiology models. Differences in model definitions fall under changes in the number of sub-populations, how transitions are computed, and time-step sizes considered. The performance gains obtained in GPU and IPU implementations are likely to transfer to all such variations of compartmental epidemiology models. More generally, this approach could provide potential benefits in several applications in life sciences, in which the challenge is to infer small set of parameters from a large-scale model which generates high-dimensional observations \cite{ProteinFoldingMonteCarlo,GWASForwardSim}.

In future, it would be also interesting to test the algorithm on the next generation
acceleration chips: the A100 from NVIDIA and the Mk2 from Graphcore.

\bibliographystyle{acm}
\bibliography{sample-base}

\end{document}